\RequirePackage{fix-cm}

\documentclass[%
 reprint,
 amsmath,amssymb,
 aps,
 prd,
 superscriptaddress,
 showkeys
]{revtex4-2}

\usepackage{graphicx}
\usepackage{dcolumn}
\usepackage{bm}
\usepackage{hyperref}
\usepackage{lipsum}
\usepackage{slantsc}
\usepackage{slashed}
\usepackage{mathtools}
\usepackage{siunitx}
\usepackage{xcolor}
\usepackage[T1]{fontenc}
\usepackage{lmodern} 
\rmfamily 

\newcommand{\inchsign}{^{\prime\prime}}

\DeclareFontShape{T1}{lmr}{b}{sc}{<->ssub*cmr/bx/sc}{}
\DeclareFontShape{T1}{lmr}{bx}{sc}{<->ssub*cmr/bx/sc}{}

\DeclarePairedDelimiterXPP\BigOSI[2]%
  {\mathcal{O}}{(}{)}{}%
  {\SI{#1}{#2}}

\makeatletter
\g@addto@macro\@floatboxreset\centering
\makeatother

\begin{document}

\preprint{APS/123-QED}

\title{Searching for Minicharged Particles at the Energy Frontier with the MoEDAL-MAPP Experiment at the LHC}

% ***** Author Info ******
\author{Vasiliki A. Mitsou}
\affiliation{Instituto de F\'isica Corpuscular, CSIC--Universitat de Val\`encia, 46980 Paterna (Val\`encia), Spain}
\affiliation{National Technical University of Athens, School of Applied Mathematical and Physical Sciences, Department of Physics, 9, Iroon Polytechniou St, Zografou Campus, 15780 Athens, Greece}

\author{Marc de Montigny}
\altaffiliation{Visiting Professor, D\'epartement de physique, de g\'enie physique et d'optique, Universit\'e Laval, Qu\'ebec, QC G1V 0A6, Canada}
\affiliation{Facult\'e Saint-Jean, University of Alberta, Edmonton, AB T6C 4G9, Canada}

\author{Abhinab Mukhopadhyay}
\affiliation{Department of Physics, Faculty of Science, University of Alberta, Edmonton, AB T6G 2E1, Canada}

\author{Pierre-Philippe A. Ouimet}
\affiliation{Department of Physics, Faculty of Science, University of Regina, Regina, SK S4S 0A2, Canada}

\author{James Pinfold}
\author{Ameir Shaa}
\affiliation{Department of Physics, Faculty of Science, University of Alberta, Edmonton, AB T6G 2E1, Canada}

\author{Michael Staelens}
\email{michael.staelens@ific.uv.es}
\affiliation{Instituto de F\'isica Corpuscular, CSIC--Universitat de Val\`encia, 46980 Paterna (Val\`encia), Spain}
\affiliation{Department of Physics, Faculty of Science, University of Alberta, Edmonton, AB T6G 2E1, Canada}

\date{\today}

% ***** Abstract ****** ---- 
\begin{abstract}
MoEDAL's Apparatus for Penetrating Particles (MAPP) Experiment is designed to expand the search for new physics at the LHC, significantly extending the physics program of the baseline MoEDAL Experiment. The Phase-1 MAPP detector (MAPP-1) is currently undergoing installation at the LHC's UA83 gallery adjacent to the LHCb/MoEDAL region at Interaction Point~8 and will begin data-taking in early 2024. The focus of the MAPP experiment is on the quest for new feebly interacting particles---avatars of new physics with extremely small Standard Model couplings, such as minicharged particles (mCPs). In this study, we present the results of a comprehensive analysis of MAPP-1's sensitivity to mCPs arising in the canonical model involving the kinetic mixing of a massless dark $U(1)$ gauge field with the Standard Model hypercharge gauge field. We focus on several dominant production mechanisms of mCPs at the LHC across the mass--mixing parameter space of interest to MAPP: Drell--Yan pair production, direct decays of heavy quarkonia and light vector mesons, and single Dalitz decays of pseudoscalar mesons. The $95\%$ confidence level background-free sensitivity of MAPP-1 for mCPs produced at the LHC's Run~3 and the HL-LHC through these mechanisms, along with projected constraints on the minicharged strongly interacting dark matter window, are reported. Our results indicate that MAPP-1 exhibits sensitivity to sizable regions of unconstrained parameter space and can probe effective charges as low as $8 \times 10^{-4}\:e$ and $6 \times 10^{-4}\:e$ for Run~3 and the HL-LHC, respectively.
\end{abstract}

% Based on JHEP keywords for [hep-ph]
                              
\maketitle

%\tableofcontents

% ***** Introduction ******
\section{\label{sec:Intro}Introduction}
While the Standard Model (SM) of particle physics is arguably one of the most successful theories we have, there now exists a strong consensus in the community that it is incomplete~\cite{Gaillard1999}. For example, it fails to adequately explain the matter--antimatter asymmetry, and it contains no viable candidate to explain the observed dark matter content of the universe. In particular, strong astrophysical evidence points to the conclusion that dark matter forms roughly $26\%$ of the mass--energy content of our universe~\cite{Planck2018cosmo}. However, to date, no direct detection of any dark matter candidates has been made.

This lack of experimental guidance has led to a proliferation of possible extensions to the SM, one of which posits that there exists a dark, or hidden, sector that minimally couples to the SM via one or more of several possible portal interactions~\cite{Patt2006,Batell2009,Beacham2020,Lanfranchi2021}. This dark sector could potentially have a rich phenomenology that is comparable, or larger in scope to that of the SM, with its own forces and matter particles that are almost completely invisible to us due to their feeble interactions with SM particles.

One observable consequence of such a dark sector would be the existence of new long-lived electrically neutral particles; another would be the existence of minicharged particles (mCPs)~\cite{Lanfranchi2021}. These mCPs are potential dark matter particles that acquire an effective electric charge that is much smaller than that of the electron due to the nature of their interactions with SM gauge fields. This second type of feebly interacting particle naturally emerges in models that include a vector portal coupling between the dark sector and the SM.

First proposed by Bob Holdom in 1986 in Refs.~\cite{Holdom1986_1, Holdom1986_2}, the vector portal couples a dark $U(1)$ gauge field to the Standard Model's $U(1)$ hypercharge gauge field in a gauge invariant way. Typically, this coupling is written as
 \begin{eqnarray}
     \frac{\kappa}{2}A'_{\mu\nu}B^{\mu\nu} , \nonumber
 \end{eqnarray}
where $A_{\mu\nu}'$ is the field strength tensor for a dark $U(1)$ gauge field, known as the dark photon, and $B^{\mu\nu}$ is the field strength tensor for $U(1)$ hypercharge. These types of couplings between two $U(1)$ gauge fields are also known as kinetic mixing terms, as they mix SM gauge fields with dark gauge fields. How such interactions induce effective visible minicharges in electrically charged dark sector matter particles will be discussed in Sec.~\ref{sec:mCPprod}.

The obvious implication of this effective electric charge, however, is that the mCPs can be thought of as coupling to the photon. This allows the use of such models to explain a number of observed discrepancies from SM predictions, such as the EDGES anomaly~\cite{Bowman2018,Munoz2018,Berlin2018,Kovetz2018,Liu2019,Aboubrahim2021,Mathur2022} and the muon anomalous magnetic moment~\cite{Bai2021,Aguillard2023}. It also opens up a large number of potential production channels for mCPs at accelerators~\cite{demontigny2023minicharged}, allowing for their direct detection.

This potential for discovery has given rise to an expansive experimental program at particle accelerators (reviewed in detail in Ref.~\cite{demontigny2023minicharged}) with several experiments both proposed and underway at the Large Hadron Collider (LHC) alone~\cite{Ball2021,Foroughi2021,Pinfold2791293,KLING2023}. In particular, the detection of mCPs is at the heart of the physics program of the MoEDAL's Apparatus for Penetrating Particles (MAPP) extension to the Monopole and Exotics Detector at the LHC (MoEDAL) experiment~\cite{StaelensPhDThesis,pinfold2023moedalmapp}. The MAPP Experiment's ground-breaking physics program encompasses numerous scenarios that offer potentially revolutionary insights into several foundational questions: what is the nature of dark matter? is there a hidden/dark sector? and what is the mechanism of electric charge quantization?

In Sec.~\ref{sec:MAPP}, we provide an update on the MAPP detector, including an important update on its location and a comprehensive overview of its design. In the subsequent sections, we will consider the potential for the discovery of mCPs at the MAPP Experiment using a typical benchmark model with massless dark photons that kinetically mix with the SM hypercharge gauge field, as was previously performed for other experiments in Refs.~\cite{Prinz1998,Haas2015,Magill2019,Kelly2019,Gninenko2019,Kim2021,Ball2021,Foroughi2021}. 

The remainder of this article is organized as follows. In Sec.~\ref{sec:mCPprod}, we address the various production channels we are considering for mCPs, including Drell--Yan pair production and production via several meson decay channels. In Sec.~\ref{sec:MAPPresults}, we discuss the sensitivity of the MAPP-1 detector, the first phase of the MAPP program, to mCPs and present projected exclusion limits for this detector. In Sec.~\ref{sec:mCSIDM}, we discuss strongly interacting dark matter models that contain mCPs and show projected sensitivity bounds for models of this type. Finally, conclusions and future outlook are provided in Sec.~\ref{sec:Conc}.

% ***** MoEDAL-MAPP ******
\section{\label{sec:MAPP}The MoEDAL-MAPP Experiment}
\begin{figure}[htb]
\includegraphics[width = 8.5 cm]{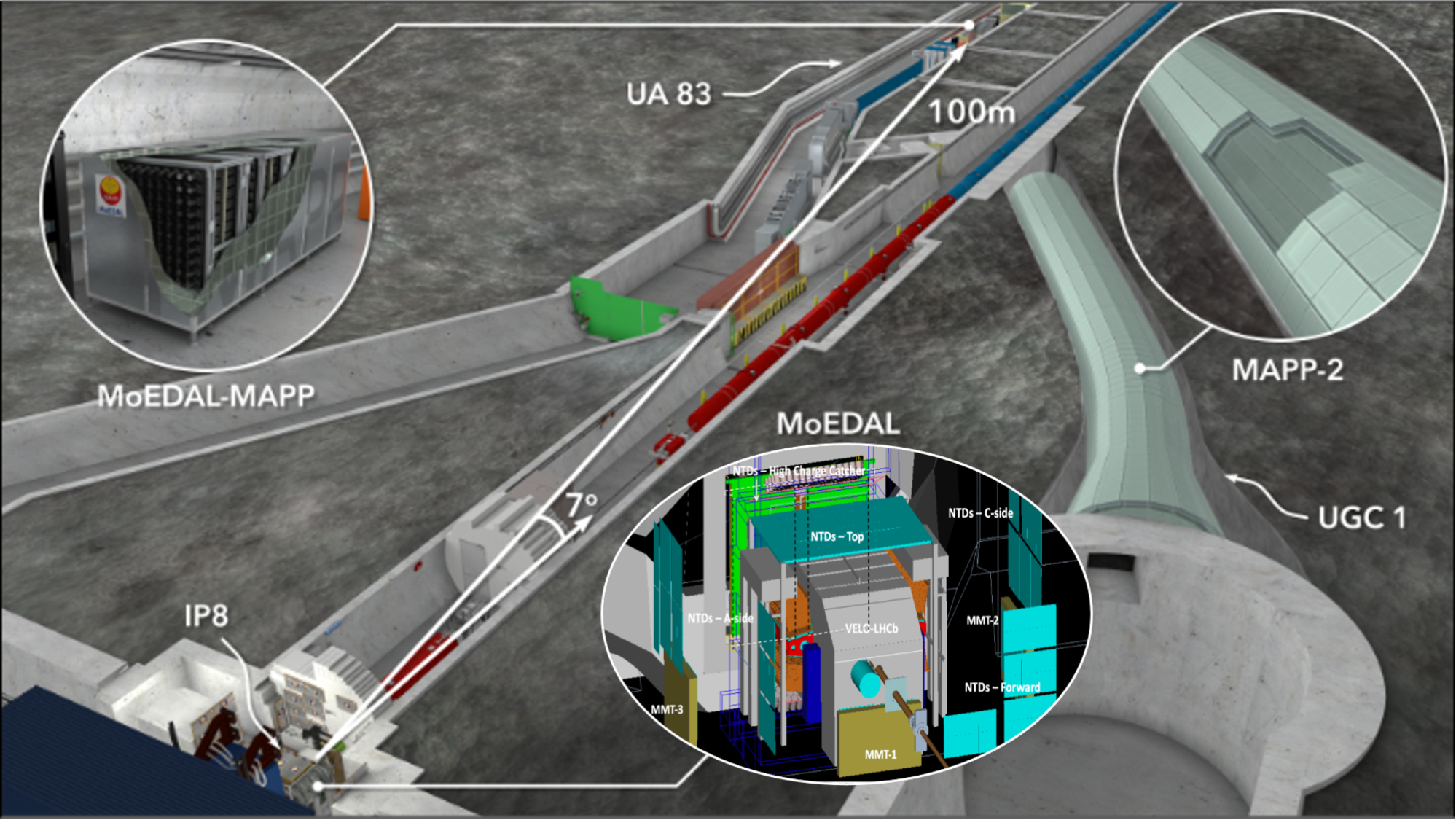}
\caption{\label{fig:fig1} An illustration depicting the arrangement of the MoEDAL-MAPP facility situated at the LHC's IP8 region. It showcases the positioning of the MoEDAL (IP8), MAPP-1 (UA83), and future MAPP-2 (UGC1) detectors.}
\end{figure}

In November 2021, the Large Hadron Collider Committee (LHCC) unanimously approved the first phase of the MAPP Experiment---a fully active scintillation detector with a sensitive volume of $\sim3$~m$^{3}$---to collect data at the LHC's Run~3~\cite{Pinfold2791293}. The combined MoEDAL-MAPP Experiment covers a wide range of beyond the Standard Model physics scenarios involving highly ionizing~\cite{DOI:10.1142/S0217751X14300506} and feebly interacting particles~\cite{Deppisch2019,FRANK2020135204,deVries2021,Dreiner2021,StaelensPhDThesis,pinfold2023moedalmapp,demontigny2023minicharged,Dreiner2023}, respectively. The Phase-1 MAPP detector (MAPP-1) is currently undergoing installation in the UA83 gallery adjacent to the MoEDAL/LHCb region at Interaction Point~8 (IP8). Based on the latest results obtained from surveying this region, the position of the center of the front face of the MAPP-1 detector is $\sim97.8$~m from the IP and at an angle of $\sim7.3^{\circ}$ with respect to the beamline. MAPP-1 has a pointing geometry, which is oriented such that it is facing toward the IP. Additionally, the UA83 gallery location benefits from a rock overburden of $110$~m, providing substantial protection from cosmic rays, as well as approximately $50$~m worth of material shielding between the detector and the IP. An overview of the MoEDAL-MAPP arena is provided in Fig.~\ref{fig:fig1}.

MAPP-1 comprises four collinear sections, each containing $100$ scintillator bar units of size $10$~cm $\times$ $10$~cm $\times$ $75$~cm coupled directly to a single high-gain ($10$-stage), low-noise $3.1\inchsign$ photomultiplier tube (PMT); such a design thus provides four-fold coincidence, reducing the rate of false signal events from dark noise in the PMTs to negligible levels. Each scintillator bar unit contains two standard blue-emitting polystyrene plastic scintillator bars of size $5$~cm $\times$ $10$~cm $\times$ $75$~cm (SP32~\cite{KAPLON2023168186} doped for enhanced light output, NUVIATech Instruments, CZ; see Appendix~B of Ref.~\cite{StaelensPhDThesis} for detailed material properties). The characterization of the scintillator bars and PMTs was conducted at the University of Alberta. The preparation of the detector materials was also performed there in the MoEDAL-MAPP Grade-C cleanroom: the scintillator bars were polished and individually wrapped in Tyvek\textsuperscript{\textregistered}, paired, wrapped in another layer of Tyvek\textsuperscript{\textregistered} followed by a layer of black paper, and then covered with two layers of black electrical tape. Silicone light guides fabricated in-house using SYLGARD™~184 silicone elastomer were then mounted to one of the ends of each of the scintillator bar units, and a PMT was attached to each light guide. Lastly, a small hole was drilled into one of the bars on the other end of each scintillator bar unit, in which a small LED was inserted and secured with optical epoxy. The MAPP-1 detector frame consists of T-slotted aluminum bars and support frames constructed from high-density polyethylene; the scintillator bar units slot into the support frames accordingly.

\begin{figure}[htb]
\includegraphics[width = 8.5 cm]{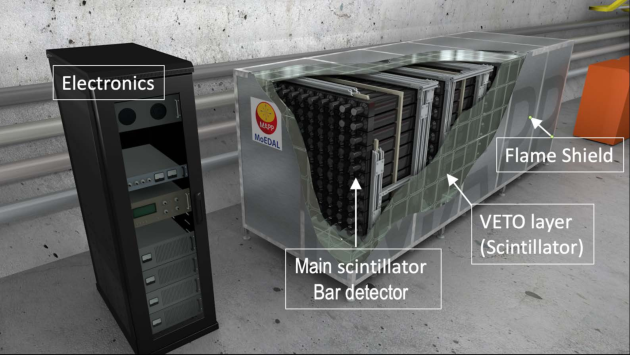}
\caption{\label{fig:fig2} A schematic of the MAPP-1 detector, emphasizing its main components.}
\end{figure}

A hermetic veto system comprising $1$~cm thick planes of scintillator surrounds the MAPP-1 detector. Each plane is assembled from $25$~cm $\times$ $25$~cm $\times$ $1$~cm scintillator subplanes containing two embedded wavelength-shifting fibers read out by silicon photomultipliers (SiPMs). Notably, due to the veto system's geometry, any through-going particle produced at the IP would yield a signal in at least two veto tiles and the detector; thus, monitoring the efficiency of the veto tiles will be possible using cosmic rays. Additionally, in the event that additional veto capabilities are required, the MAPP-1 detector can be fiducialized, i.e., the outer layer of scintillator bars in each section of the detector can be utilized as an additional veto, at the expense of reducing the sensitive volume of the detector (which would then comprise a total of $256$ scintillator bar units---$64$ per section). Lastly, the entire detector is completely encased in an aluminum flame shield. All of the readout electronics are located in the UA83 gallery behind the detector. A schematic of the MAPP-1 detector is provided in Fig.~\ref{fig:fig2}. 

\textit{In situ} calibration of the MAPP-1 detector will be performed using two methods: 1) calibration employing the built-in system of LEDs embedded in the scintillator bars, and 2) calibration using incident muons produced at the IP that pass through all four collinear sections of the detector. For maximum flexibility, a software trigger scheme will be followed as far as possible. However, a number of software trigger categories will also be predefined. An important trigger preset relating to mCP detection requires quadruple-coincident photoelectron production from a through-going mCP with sufficient energy deposition in each of the four collinear sections. Data collected by MAPP will be stored locally on a data server situated in the UA83 gallery, which will be connected via Ethernet. The data will be sent to several off-site PCs with redundant data storage capabilities for backup storage and data analysis purposes.

% ***** mCP Production at LHC ******
\section{\label{sec:mCPprod}Minicharged Particle Production at the LHC}
\begin{figure}[htb]
\includegraphics[width = 9 cm]{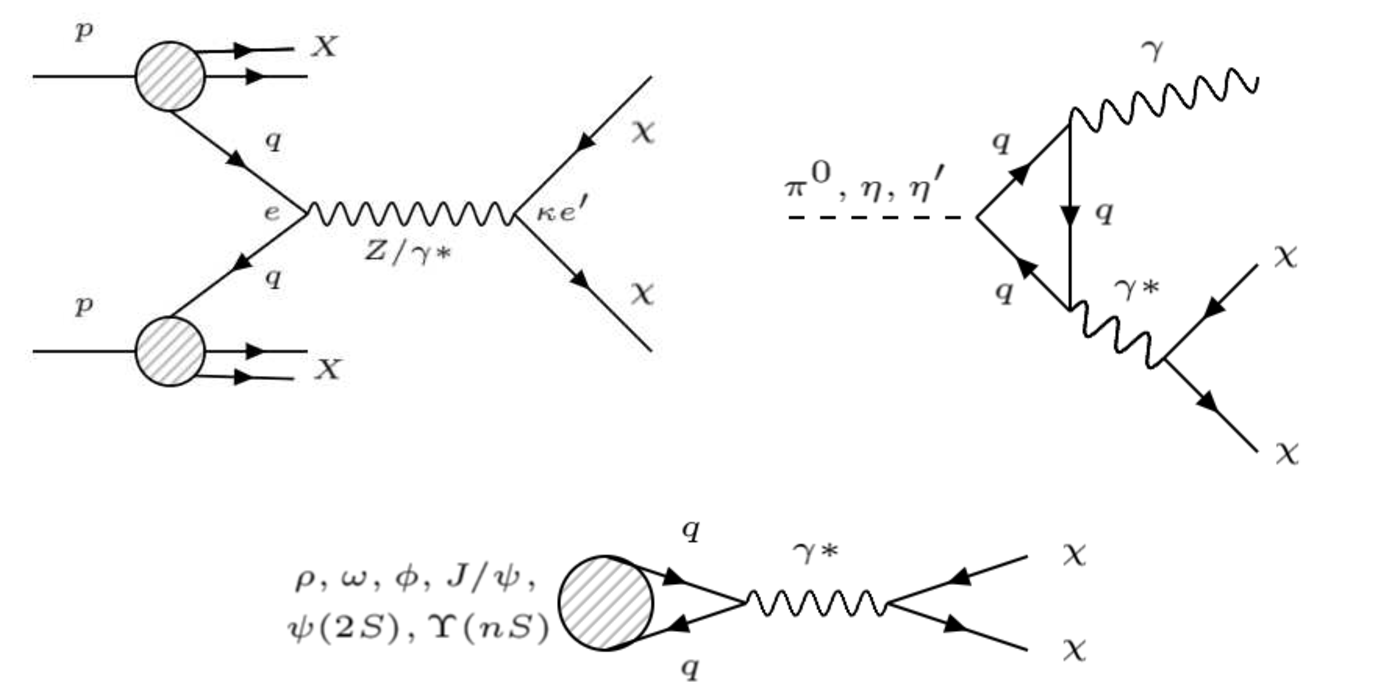}
\caption{\label{fig:fig3} Feynman diagrams of the various minicharged particle production modes considered in this study.}
\end{figure}

Fermionic minicharged particles can be copiously produced at the LHC through a variety of mechanisms, such as the Drell--Yan mechanism, meson decays, and proton bremsstrahlung. In this study, we investigated the sensitivity of the MAPP-1 experiment at the LHC's Run~3 and the HL-LHC for mCPs produced via the Drell--Yan mechanism, direct decays of heavy quarkonia and light vector mesons, as well as single Dalitz decays of pseudoscalar mesons. Representative Feynman diagrams for each of these processes are shown in Fig.~\ref{fig:fig3}.

\subsection{\label{subsec:DYprod}Drell--Yan Pair Production}
We developed a \textsc{FeynRules}~\cite{CHRISTENSEN2009,Christensen2011} model to simulate the production of mCPs via the Drell--Yan process with \textsc{MadGraph5}\_aMC@NLO (MG5; v2.7.3)~\cite{Alwall2011,Alwall2014} based on Holdom's original model that predicts such particles~\cite{Holdom1986_1}. The model adds to the SM an additional massless $U(1)$ gauge field $A'_{\mu}$, hereafter referred to as the dark photon, that interacts with the SM hypercharge gauge field $B^{\mu}$ through a kinetic mixing interaction. A new massive Dirac fermion ($\chi$) coupled to the dark photon gauge field and thus charged under it with an electric charge, $e'$, is also included. Generally, the Lagrangian can be written as
\begin{align}
\mathcal{L} & = \mathcal{L}_{\mathrm{SM}} -\frac{1}{4}  A'_{\mu \nu} A'^{\mu \nu}  + i \bar{\chi} \left( \slashed{\partial} + i e' \slashed{A}' +i m_{\chi} \right) \chi \nonumber \\
& - \frac{\kappa}{2} A'_{\mu \nu} B^{\mu \nu}, \label{EQN:Lag1}
\end{align}
where $m_{\chi}$ is the dark fermion mass, $\kappa$ is an arbitrary (and potentially irrational) small parameter that controls the strength of the kinetic mixing, and $A'_{\mu \nu}$ is the dark photon field strength tensor following the usual definition of $A'_{\mu \nu} = \partial_{\mu} A'_{\nu} - \partial_{\nu} A'_{\mu} $. 

The non-diagonal kinetic mixing term in the Lagrangian in Eq.~\ref{EQN:Lag1} can be diagonalized through a field redefinition, $A'_{\mu} \Rightarrow A'_{\mu} + \kappa B_{\mu}$~\cite{Haas2015}. Applying this field redefinition, we arrive at the following Lagrangian,
\begin{align}
\mathcal{L} &  = \mathcal{L}_{\mathrm{SM}} -\frac{1}{4}  A'_{\mu \nu} A'^{\mu \nu}  \nonumber \\
& + i \bar{\chi} \left(   \slashed{\partial} + i e' \slashed{A}' - i \kappa e' \slashed{B} + i m_{\chi} \right) \chi, \label{EQN:Lag2}
\end{align}
which uncovers a coupling between the SM hypercharge gauge field and the charged matter field $\chi$. The new fermionic field $\chi$ thus behaves as a field charged under SM hypercharge with an electric charge of $\kappa e'$, i.e., appearing minicharged in the visible sector. Moreover, $\chi$ couples to the photon and $Z^{0}$ boson with couplings of $\kappa e' \cos{\theta_{\mathrm{W}}}$ and $-\kappa e' \sin{\theta_{\mathrm{W}}}$, respectively. Expressing its effective charge in terms of the elementary charge thus yields $\epsilon \equiv \kappa e' \cos{\theta_{\mathrm{W}}}/e$. Under the assumption that the new $U(1)$ gauge symmetry remains unbroken, the new charged matter field is stable.

Following Ref.~\cite{Haas2015}, the new gauge fields, parameters, and Lagrangian (Eq.~\ref{EQN:Lag2}) of the model were implemented into a \textsc{FeynRules} file that was subsequently used to generate a \textsl{Universal \textsc{FeynRules} Output} (UFO)~\cite{DEGRANDE2012} model to be read by MG5. For pair production via the Drell--Yan mechanism, the number of mCPs $\left( N_{\chi} \right)$ produced at a particular LHC run can be estimated as follows,
\begin{equation}
    N_{\chi} \simeq 2 \sigma_{q\bar{q} \rightarrow \chi \bar{\chi}} L^{\mathrm{int}}_{\mathrm{LHCb}},  \label{EQN:NmCP_DY}
\end{equation}
where $\sigma_{q\bar{q} \rightarrow \chi \bar{\chi}}$ represents the Drell--Yan pair-production cross-section, and $L^{\mathrm{int}}_{\mathrm{LHCb}}$ is the estimated integrated luminosity at IP8 for a given LHC run, assumed to be $30$ and $300$~fb$^{-1}$ for Run~3 and the HL-LHC, respectively.

\subsection{\label{subsec:Mdec}Meson Decays}
Through their coupling to the photon, mCPs can be abundantly produced through a variety of meson decays. Similar to Eq.~\ref{EQN:NmCP_DY}, the approximate number of mCPs produced via meson decays can be estimated from
\begin{equation}
    N_{\chi} \simeq 2   \mathcal{B}_{M\rightarrow \chi \bar{\chi} X } \sigma_{M}  L^{\mathrm{int}}_{\mathrm{LHCb}},   \label{EQN:NmCP_MDec}
\end{equation}
where $\mathcal{B}_{M\rightarrow \chi \bar{\chi} X }$ and $\sigma_{M}$ are the branching ratio to mCPs and the production cross-section of the particular meson under consideration, respectively. The calculations of $\mathcal{B}_{M\rightarrow \chi \bar{\chi} X }$ for the direct decays and Dalitz decays under consideration are detailed in turn in the following two subsections.

\subsubsection{\label{subsec:Dir}Direct Decays of Vector Mesons}
Neutral vector mesons (e.g., $\rho$, $\omega$, $\phi$, $J/\psi$, $\psi\left(2S\right)$, $\Upsilon\left(nS\right)$) can decay electromagnetically to a pair of mCPs directly. The branching ratio for these decays, $\mathcal{B}_{M\rightarrow \chi \bar{\chi}}$, can be calculated by rescaling the corresponding SM branching ratio to an electron--positron pair ($\mathcal{B}_{M \rightarrow e^{-} e^{+}}$) by $\epsilon^{2}$ and a phase-space factor associated with the mass of the mCP~\cite{Kelly2019}. Explicitly, the branching ratio for direct decays of neutral vector mesons to mCPs can be written as
\begin{equation}
    \mathcal{B}_{M\rightarrow \chi \bar{\chi}} = \epsilon^{2} \mathcal{B}_{M \rightarrow e^{-} e^{+}} I^{(2)} \left(\frac{m_{\chi}^2}{m_{M}^{2}},\frac{m_{e}^2}{m_{M}^{2}}\right),   \label{EQN:BRdir1}
\end{equation}
where $m_{M}$ and $m_{e}$ denote the meson and electron masses, respectively, and $I^{(2)}\left(x,y\right)$ is a function that describes the two-body decay~\cite{Kelly2019},
\begin{equation}
    I^{(2)}\left(x,y\right) = \frac{\left(1+2x\right)\sqrt{1-4x}}{\left(1+2y\right)\sqrt{1-4y}}.     \label{EQN:I2}
\end{equation}
Thus, the following formula for the branching ratio of direct decays of neutral vector mesons to mCPs is obtained,
\begin{equation}
    \mathcal{B}_{M\rightarrow \chi \bar{\chi}} = \epsilon^{2} \mathcal{B}_{M \rightarrow e^{-} e^{+}} \frac{\left( m_{M}^{2} + 2m_{\chi}^{2} \right) \sqrt{m_{M}^{2}-4m_{\chi}^2}}{\left(m_{M}^{2} + 2m_{e}^{2} \right) \sqrt{m_{M}^{2} - 4m_{e}^{2}}}.   \label{EQN:BRdir2}
\end{equation}

\subsubsection{\label{subsec:Dal}Dalitz Decays of Pseudoscalar Mesons}
Neutral pseudoscalar mesons (e.g., $\pi^0$, $\eta$, $\eta'$) can decay to a photon and a pair of mCPs through single Dalitz decays. In this study, we consider only single Dalitz decays since, although double Dalitz decays to two pairs of mCPs are possible in principle, they are highly suppressed by $\epsilon^{4}$. The branching ratio for single Dalitz decays of neutral pseudoscalar mesons to mCPs can be calculated as follows,
\begin{equation}
    \mathcal{B}_{M\rightarrow \gamma \chi \bar{\chi}} = \epsilon^{2} \alpha \mathcal{B}_{M \rightarrow \gamma \gamma} I^{(3)}\left(\frac{m_{\chi}^{2}}{m_{M}^{2}} \right),
\end{equation}
where $\mathcal{B}_{M \rightarrow \gamma \gamma}$ denotes the corresponding SM branching ratio to two photons, and $I^{(3)}(x)$ describes the three-body decay~\cite{Kelly2019},
\begin{equation}
    I^{(3)}(x) = \frac{2}{3\pi} \int_{4x}^{1} dz \sqrt{1 - \frac{4x}{z}} \frac{\left(1-z\right)^{3}\left(2x+z\right)}{z^2}.
\end{equation}
Thus, the following formula for the branching ratio of single Dalitz decays of neutral pseudoscalar mesons to mCPs is obtained,
\begin{widetext}
\begin{equation}
\label{EQN:BRDal}
    \mathcal{B}_{M\rightarrow \gamma \chi \bar{\chi}} = \epsilon^{2} \alpha \mathcal{B}_{M \rightarrow \gamma \gamma} \left(\frac{2}{3\pi} \int_{4m_{\chi}^{2}/m_{M}^{2}}^{1} dz \sqrt{1 - \frac{4m_{\chi}^{2}/m_{M}^{2}}{z}} \frac{\left(1-z\right)^{3}\left(2m_{\chi}^{2}/m_{M}^{2}+z\right)}{z^2}\right).
\end{equation}
\end{widetext}

\begin{figure}[htb]
\includegraphics[width = 8 cm, height = 6 cm]{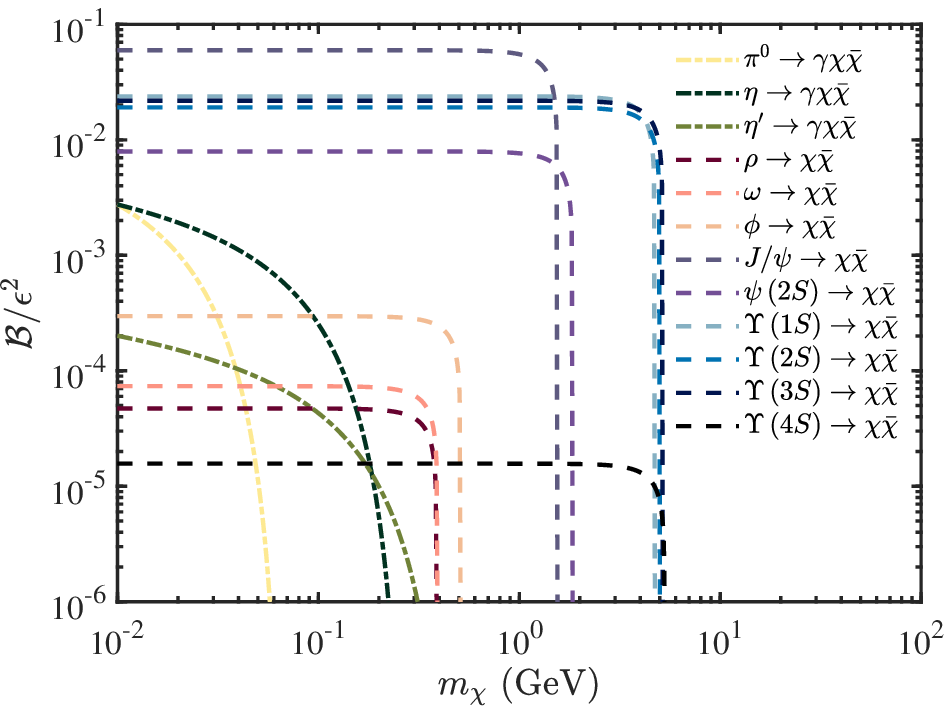}
\caption{\label{fig:fig4} Calculated branching ratios $\left(\mathcal{B}\right)$ for the leading-order decays of various pseudoscalar mesons, light vector mesons, and heavy quarkonia to a pair of minicharged particles normalized by $\epsilon^2$. The corresponding results for Dalitz decays and direct decays are denoted by dash-dotted and dashed lines, respectively.}
\end{figure}

In our calculations of Eqs.~\ref{EQN:BRdir2} and \ref{EQN:BRDal} for each of the decays studied, the corresponding meson masses ($m_{M}$) and SM branching ratios ($\mathcal{B}_{M \rightarrow e^{-} e^{+}}$, $\mathcal{B}_{M \rightarrow \gamma \gamma}$) were taken directly from the 2022 PDG review~\cite{PDG2022}. For Eq.~\ref{EQN:BRDal}, we used the nominal value of the fine-structure constant, $\alpha = 1/137$, and performed the integrations numerically with \textit{Mathematica} (v11.2). Additionally, both equations and their respective calculations were quickly verified by setting $\epsilon=1$ and $m_{\chi}=m_{e}$ and ensuring that the correct SM branching ratio values were recovered for each meson decay channel studied. Branching ratios calculated for the direct decays of heavy quarkonia and light vector mesons and single Dalitz decays of pseudoscalar mesons to mCPs, normalized by $\epsilon^{2}$, are presented in Fig.~\ref{fig:fig4}.

\subsection{\label{subsec:ProdXS}Production Cross-Sections and Validation}
Simulation-based estimates were performed to obtain the mCP production cross-sections for each of the processes described in the previous section. In the case of Drell--Yan pair-produced mCPs, leading-order cross-sections were calculated with our MG5 model, which had previously undergone thorough testing and validation; these results can be found in Ref.~\cite{StaelensPhDThesis}. All of our MG5 simulations were performed using the default parton distribution function (NNPDF2.3QED~\cite{Ball2013}) and the approximate $Z$-pole value of the fine-structure constant, $\alpha\left(m_{Z}\right) \simeq 1/127.94$~\cite{PDG2022}. For direct decays of heavy quarkonia to mCPs, we estimated $\sigma_{M}$ using cross-sections and multiplicities extracted from $pp$ collisions simulated with \textsc{Pythia}~8 (v8.240)~\cite{SJOSTRAND2015159,Bierlich2022} using the default Monash 2013 tune~\cite{Skands2014}. In the case of the pseudoscalar and light vector mesons, we normalized the cross-sections based on the inelastic $pp$ cross-section, estimated as $79.57$~mb and $79.95$~mb for Run~3 ($\sqrt{s}=13.6$~TeV) and the HL-LHC ($\sqrt{s}=14$~TeV), respectively~\cite{Aaboud2016}.

\begin{figure}[htb]
\includegraphics[width = 8 cm, height = 6 cm]{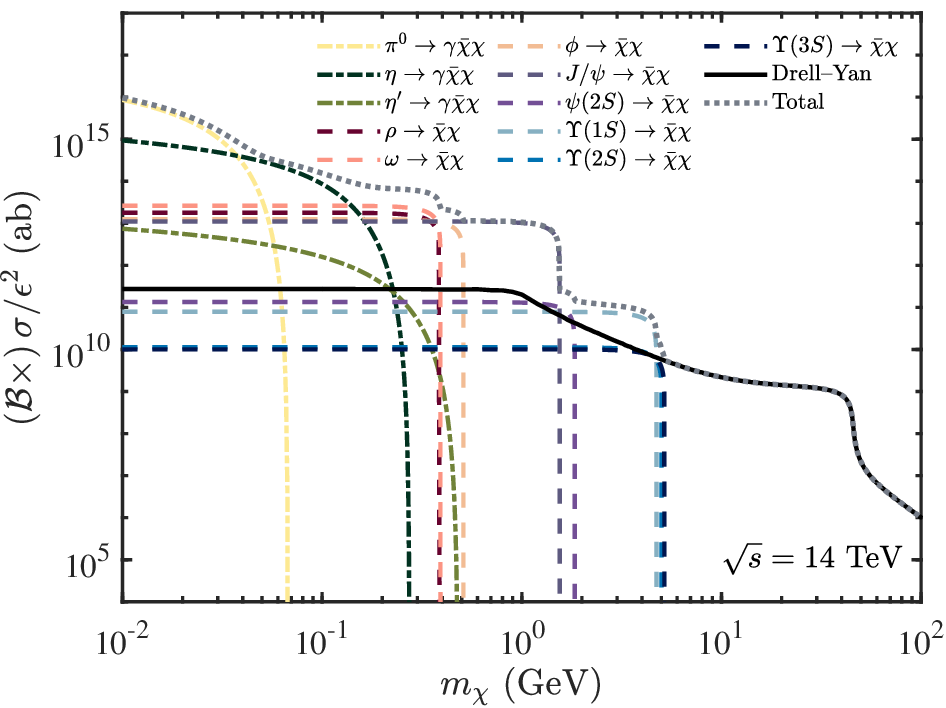}
\caption{\label{fig:fig5} Charge-normalized minicharged particle production cross-sections at $\sqrt{s}=14$~TeV for all processes considered in this study. Results corresponding to meson decays are scaled by their respective branching ratios.}
\end{figure}

A comparison with the literature was performed to validate our mCP production cross-section results for meson decays. In particular, we compared with the $\sqrt{s}=13$~TeV results published by the milliQan Collaboration in their recent search for mCPs at the LHC's Run~2~\cite{Ball2020}, which were computed using a mixture of simulations, theoretical calculations, and available measurements from past LHC runs. Our estimates are largely in agreement with their results. In Fig.~\ref{fig:fig5}, we present our results for all the production mechanisms considered at a center-of-mass energy of $\sqrt{s}=14$~TeV, with the total cross-section overlayed.

% ***** Sensitivty of MAPP to mCPs ******
\section{\label{sec:MAPPresults}Sensitivity of MAPP-1 to Minicharged Particles}
The sensitivity of the MAPP-1 detector to mCPs over the mass--mixing parameter space accessible at the upcoming LHC runs was determined based on the following formula for the estimated number of signal events ($N_{\mathrm{sig}}$) detected:
\begin{equation}\label{Eqn:Nsig}
    N_{\mathrm{sig}} = N_{\chi} \times A \times P
\end{equation}
where $N_{\chi}$ is estimated for a particular production mechanism using Eq.~\ref{EQN:NmCP_DY} or \ref{EQN:NmCP_MDec}; $A$ represents the acceptance of the detector to mCPs produced by such a process; and $P$ represents the probability of detecting an mCP, which follows a Poisson distribution associated with the number of photoelectrons produced in the detector by a through-going mCP. Monte Carlo event generator simulations were performed for all of the production mechanisms described in Sec.~\ref{sec:mCPprod} to obtain estimates of the geometric acceptance of the MAPP-1 detector, and acceptance tables were constructed from the results. In particular, for Drell--Yan pair-produced mCPs in $pp$ collisions at center-of-mass energies of both $13.6$~TeV and $14$~TeV, $5$ million events were simulated per value of mCP mass studied using our MG5 model with the default PDF. Additionally, meson samples comprising either $2.5$ million events for heavy quarkonia or $1$ million events for pseudoscalar and light vector mesons were generated with \textsc{Pythia}~8 and decayed exclusively to mCPs. Following Ref.~\cite{Foroughi2021}, we include the \texttt{SuppressSmallPT} user hook in \textsc{Pythia}~8 to suppress the overproduction of heavy quarkonia at low $p_{\mathrm{T}}$. Specifically, this includes a suppression factor of
\begin{equation}\label{Eqn:pTsupp}
    \frac{p_{\mathrm{T}}^{4}}{\left(\left(k p_{\mathrm{T}0} \right)^{2} + p_{\mathrm{T}}^{2} \right)^{2}} \left( \frac{ \alpha_{\mathrm{s}}\left( \left(  k p_{\mathrm{T}0} \right)^2 + Q^2_{\mathrm{ren}} \right)}{\alpha_{\mathrm{s}}\left( Q^2_{\mathrm{ren}} \right)} \right)^{n} , \nonumber
\end{equation}
where $p_{\mathrm{T}}$ and $p_{\mathrm{T}0}$ are the transverse momentum and energy-dependent dampening scale, respectively, and $Q_{\mathrm{ren}}$ is the renormalization scale. We selected $k = 0.35$ and $n = 3$ based on the results reported by the authors of Ref.~\cite{Foroughi2021}, who confirmed consistency between the simulated results and measurements at LHCb for these particular parameter values. Subsequently, all sets of events generated were analyzed to determine the number of mCPs accepted by the MAPP-1 detector, where, in this case, ``accepted'' was defined as an mCP with a trajectory such that it traverses all four collinear sections of the MAPP-1 detector entirely.

For a scintillation detector with $n$ layers, the detection probability for a through-going charged particle is given by $P=\left(1 - {\rm e}^{-N_{\mathrm{PE}}}  \right)^{n}$, where $N_{\mathrm{PE}}$ represents the number of photoelectrons detected. In general, $N_{\mathrm{PE}}$ is proportional to the number of optical scintillation photons reaching the PMT ($N_{\gamma}$) and its quantum efficiency ($\mathrm{QE}$). For an mCP, $N_{\mathrm{PE}}$ can be approximated by determining $N_{\gamma}$ for a minimally ionizing charged particle and scaling the result by $\epsilon^2$, i.e., $N_{\mathrm{PE}} \propto \epsilon^2 N_{\gamma} \mathrm{QE}$. For the XP72B22 PMTs utilized in the MAPP-1 detector, we consider a $\mathrm{QE}$ of $25\%$ in our estimates, which is consistent with both the manufacturer specifications and the measurements reported by the JUNO experiment in their characterization tests of these PMT modules~\cite{CAO2021}. We estimate $N_{\gamma}$ by performing \textsc{Geant4}~\cite{AGOSTINELLI2003} (v10.6.p02) simulations to study the energy losses of minimally ionizing incident muons passing through a MAPP-1 scintillator bar unit. The scintillator bar properties and dimensions implemented were consistent with those outlined in Sec.~\ref{sec:MAPP}, including simple Tyvek\textsuperscript{\textregistered} wrapping (from default \textsc{Geant4} libraries) and assuming an overall surface reflectivity of $98\%$, a bulk light attenuation length of $2.6$~m, and a light output of $10000$ photons/MeV. Additionally, we modeled the silicone light guide at the end of the scintillator bar based on the material properties of silicone and assuming a refractive index of $1.44$~\cite{CAI2010274,Qiu2012}. Optical photon transport in \textsc{Geant4} was performed using the UNIFIED~\cite{Nayar1991,Levin1996} and LUT Davis models~\cite{Roncali2013,Roncali2017}. The small, but not insignificant, number of Cherenkov photons was ignored in this analysis. For $100$ events involving incident muons with a kinetic energy of $1$~GeV, the average count of optical photons reaching the PMT obtained is approximately $N_{\gamma} \simeq 6.824 \times 10^5$. Thus, we approximate the number of photoelectrons produced in a MAPP-1 scintillator bar by a through-going mCP as $N_{\mathrm{PE}} = 1.706 \times 10^5 \epsilon^2$.

In this study, projected exclusion limits were estimated under a background (BG)-free assumption. MAPP-1 is protected from cosmic ray BGs by an overburden of $110$~m of rock, from collision-related BGs by approximately $50$~m of rock, and from low energy beam BGs by $8$~m of concrete. The MAPP-1 detector will also take advantage of several BG rejection techniques by utilizing its hermetic veto system, as well as quadruple-coincidence requirements for the signal.

Despite being protected, cosmic ray muons and subsequently generated showers of gamma rays and electrons in the material overburden; a residual flux of neutrons, muons, and gamma rays expected from the beam; and a low rate of incident high-energy muons produced in the collisions, which can also produce secondaries in the detector, comprise the dominant sources of BGs expected at MAPP. This flux of muons is also used in the calibration of the MAPP-1 detector. The beam-related backgrounds in the UA83 tunnel have been studied previously by the LHC machine group using FLUktuierende KAskade (FLUKA~\cite{Ferrari2005,Bohlen2014}); the results of this study can be found in Ref.~\cite{Pinfold2791293}. Additionally, the \textsc{Geant4} simulation, SUMMA (SimUlation of the MoEDAL-MAPP Arena)~\cite{SUMMA}, was developed to perform comprehensive studies of the remaining backgrounds, which are currently in progress. Beam-off BG measurements and beam-on BG measurements that take advantage of the bunch spacing in the LHC will be conducted to characterize the cosmic ray BGs and beam- and collision-related BGs, respectively.

Dark current in the PMTs can also result in signal-like events; however, the four-fold coincidence design employed essentially eliminates this BG source, as demonstrated in the following estimation of the expected random coincidence rate ($R$) in MAPP-1. For a detector with $n$ layers, the random coincidence rate can be estimated as $R = n (\mathrm{DCR})^n \tau^{n-1}$, where $\mathrm{DCR}$ is the dark count rate, and $\tau$ is the coincidence time window (between layers). We assume a mean $\mathrm{DCR}$ of $500$~Hz, which is in accordance with the manufacturer's specifications and the measurements reported by the JUNO experiment~\cite{CAO2021} (at $0.25$~PE) and consistent with our estimates (and those reported in Ref.~\cite{Haas2015}) based on the standard dark-count spectrum presented in Fig.~65 of Ref.~\cite{burle1980PMT}. For a timing window of $\tau=15$~ns, we obtain a random coincidence rate of $R=8.4375 \times 10^{-13}$~s$^{-1}$ per coincidence path. Since the MAPP-1 detector sections each comprise a $10\times10$ array of scintillator bars, i.e., $100$ possible coincidence paths, the overall random coincidence rate in MAPP-1 is $8.4375 \times 10^{-11}$~s$^{-1}$. Assuming a $3$-year trigger live time, a negligible result of $0.008$ total signal-like events is obtained.

\begin{figure}[htb]
\includegraphics[width = 8 cm, height = 6 cm]{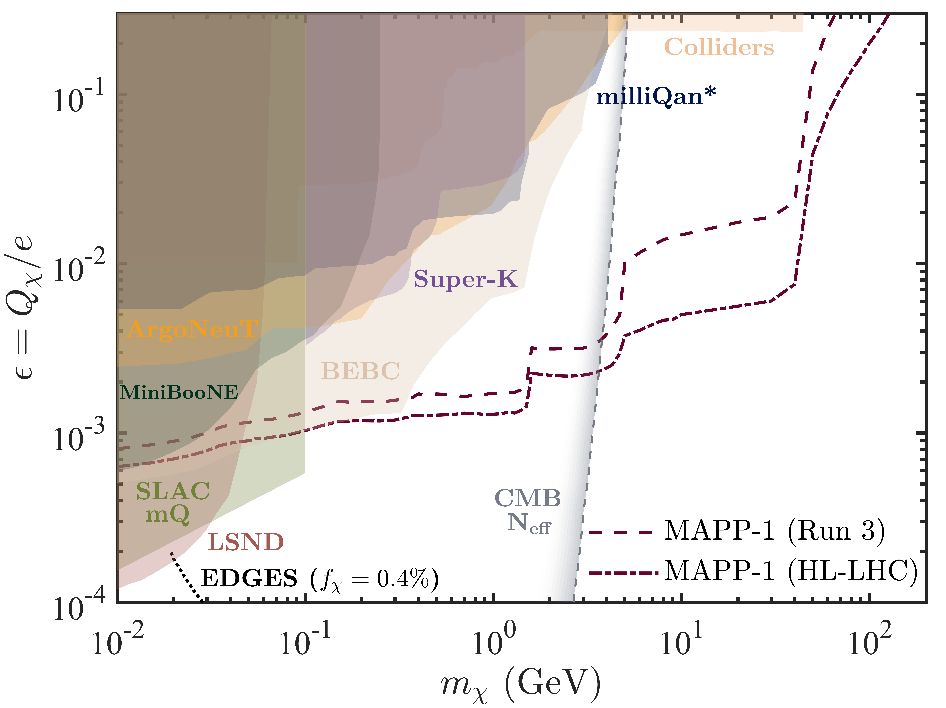}
\caption{\label{fig:fig6} The projected $95\%$ confidence level (CL) exclusion limits for MAPP-1 for minicharged particles produced in $pp$ collisions via the Drell--Yan mechanism, direct decays of heavy quarkonia and light vector mesons, and single Dalitz decays of pseudoscalar mesons: Run~3---$\sqrt{s}=13.6$~TeV, $L^{\mathrm{int}}_{\mathrm{LHCb}} = 30$~fb$^{-1}$; HL-LHC---$\sqrt{s}=14$~TeV, $L^{\mathrm{int}}_{\mathrm{LHCb}} = 300$~fb$^{-1}$. The shaded areas represent the regions of parameter space excluded by previous searches at the $95\%$ CL (except the BEBC bounds, which correspond to the $90\%$ CL)~\cite{Prinz1998,Davidson2000,Magill2019,Acciarri2020,Ball2020,Plestid2020,Marocco2021}. The grey dashed and black dotted lines represent the most stringent indirect $2\sigma$ upper limits derived from the Planck full-mission results~\cite{Planck2018cosmo} on the effective number of different neutrino species ($N_{\mathrm{eff}}$)~\cite{Adshead_2022} and the region associated with a potential resolution of the EDGES anomaly ($f_{\chi} = 0.4\%$)~\cite{Kovetz2018}, respectively.}
\end{figure}

In the $CL_{s}$ method~\cite{Read2002}, under a BG-free assumption, $95\%$ confidence level (CL) exclusion limits correspond to the region of the mass--mixing parameter space that predicts $3$ or more detected signal events~\cite{PDG2022}. Thus, for each value of mCP mass studied over the parameter space of interest, we solved Eq.~\ref{Eqn:Nsig} numerically for the corresponding value of $\epsilon$ that yields $N_{\mathrm{sig}} = 3$. The resulting $95\%$~CL exclusion limits for Run~3 and the HL-LHC assuming $30$ and $300$~fb$^{-1}$, respectively, are presented in Fig.~\ref{fig:fig6}. All the previous searches constraining this region of the mass--mixing parameter space~\cite{Prinz1998,Davidson2000,Magill2019,Acciarri2020,Ball2020,Plestid2020,Marocco2021} and the latest indirect $2\sigma$ upper limits based on the effective number of different neutrino species ($N_{\mathrm{eff}}$) determined from precise measurements of the CMB~\cite{Adshead_2022} are featured. The region associated with an mCP-based resolution of the EDGES anomaly, assuming the maximal minicharged fraction of dark matter $\left(f_{\chi}=0.4\%\right)$ consistent with constraints from Planck~2015 CMB data~\cite{Planck2015cosmo,Planck2015cosmoII}, is also shown~\cite{Kovetz2018}.

For the LHC's Run~3, the projected exclusion limits for MAPP-1 probe new parameter space for mCP masses in the range of $\sim0.35$--$70$~GeV. At the HL-LHC, assuming a $10$-fold increase in integrated luminosity, MAPP-1 exhibits sensitivity at the sub-millicharge level for a sizable region of parameter space and covers unexplored parameter space for mCP masses of $0.2 \lesssim m_{\chi} \lesssim 130$~GeV. In previous work~\cite{demontigny2023minicharged}, we established Drell--Yan-only exclusion limits assuming a $100\%$ detector efficiency; here, we significantly improved those limits across the entire mass range in which the meson decays studied contribute, even after incorporating signal efficiency estimates. Notably, the projected sensitivity of MAPP-1 is highly competitive with the bounds reported by the other two dedicated mCP search experiments at the LHC---milliQan~\cite{Ball2021} and the FORward MicrOcharge SeArch (FORMOSA)~\cite{Foroughi2021}---as well as other approved experiments, such as the SUB-Millicharge ExperimenT (SUBMET) experiment at J-PARC~\cite{Kim2021}.

% ***** Sensitivty of MAPP to mC-SIDM ******
\section{\label{sec:mCSIDM}Minicharged Strongly Interacting Dark Matter}
Traditionally, searches for dark matter have primarily focused on the weak-interaction regime, employing specific detection techniques designed to target particles that interact weakly with ordinary matter. Weakly interacting massive particles (WIMPs), the most well-known example of such particles, have been a focal point in dark matter search experiments~\cite{Roszkowski2018}. Typically, large detector volumes comprising materials that offer high sensitivity to low-energy nuclear recoils, such as liquid xenon and argon, are deployed deep underground to conduct direct-detection searches in an environment with reduced cosmic rays and background radiation. A myriad of such search experiments have been conducted, resulting in very strong constraints on the WIMP parameter space~\cite{Undagoitia2016}. Researchers have thus expanded the search to include other well-motivated dark matter candidates, such as axions~\cite{Peccei1977,Weinberg1978,Wilczek1978} and strongly interacting dark matter (SIDM)~\cite{Starkman1990,McGuire2001}. The latter models consider a scenario in which dark matter comprises particles that interact with SM particles with cross-sections above those associated with weak interactions (i.e., it does not refer to interactions mediated by the strong force). Terrestrial direct-detection experiments are limited in their ability to search for such particles as there are critical cross-section values above which the particles interact too strongly and lose the majority of their energy in the atmosphere and geosphere before they reach the detectors~\cite{Emken2018}.

Although the possibility of dark matter comprised entirely of mCPs is very strongly constrained (see, e.g., Ref.~\cite{PandaX2023}), as mentioned in Sec.~\ref{sec:MAPPresults}, a small minicharged subcomponent of dark matter of $f_{\chi} \leq 0.4$\% remains consistent with CMB data and could help to explain a portion of the dark matter abundance. This scenario of minicharged SIDM (mC-SIDM) has been discussed throughout the literature~\cite{Mahdawi2018,Emken2019,Plestid2020,Foroughi2021,Alexander2021,KLING2023}. Additionally, various recent studies have recognized an unexplored region of parameter space~\cite{Mahdawi2018,Emken2019,Plestid2020}---the so-called minicharged SIDM window---a large subregion of which current accelerator-based experiments such as MAPP could probe. 

Limits in the mC-SIDM scenario are typically presented in terms of a reference cross-section for DM--electron scattering~\cite{Essig2012,Emken2019,Foroughi2021}, 
\begin{equation}\label{Eqn:RefXS}
    \bar{\sigma}_{e\mathrm{,ref}} = 16 \pi \alpha^{2} \epsilon^{2} \mu^{2}_{\chi e}/q^{4}_{d\mathrm{,ref}}.
\end{equation}
Here, $\mu_{\chi e}$ represents the reduced mass of $\chi$ and the electron, and $q_{d\mathrm{,ref}}$ denotes the reference 3-momentum transfer. Following Refs.~\cite{Essig2012,Emken2019,Foroughi2021}, $q_{d\mathrm{,ref}} = \alpha m_{e}$ is selected based on the standard momentum transfer in DM--electron interactions for semiconductor and noble liquid materials. 

\begin{figure}[htb]
\includegraphics[width = 8 cm, height = 6 cm]{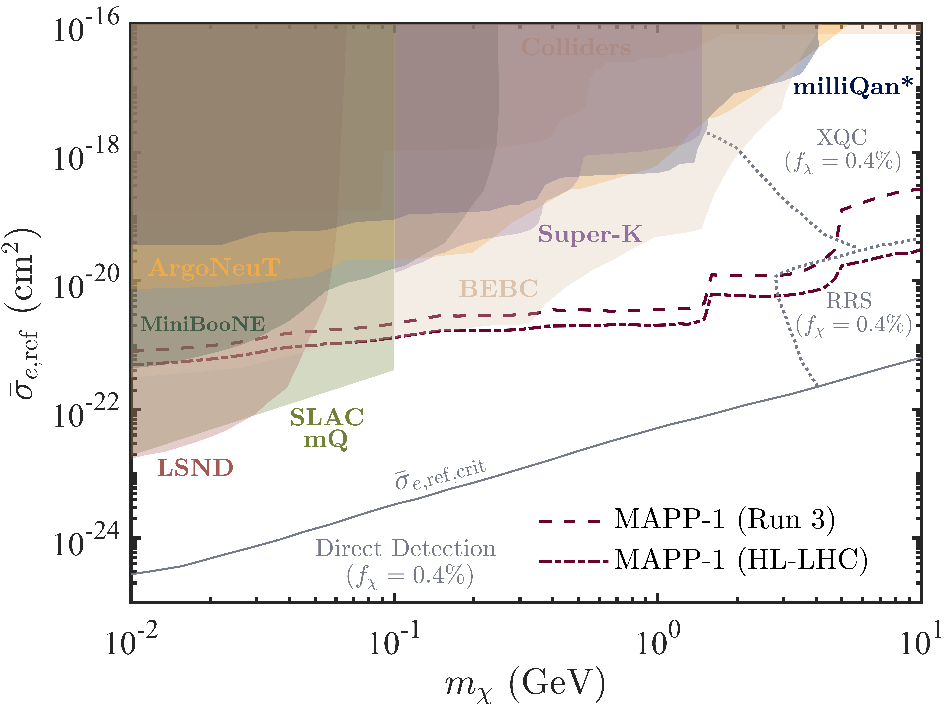}
\caption{\label{fig:fig7} The sensitivity of MAPP-1 to minicharged strongly interacting dark matter at the LHC's Run~3 and the HL-LHC established at the $95\%$ confidence level. Additional exclusion limits shown on the plot in grey correspond to projections of constraints set by several strongly interacting dark matter searches assuming a small minicharged fraction of dark matter of $f_{\chi} = 0.4$\%: direct-detection experiments~\cite{Mahdawi2018,Emken2019}, the X-ray quantum calorimetry (XQC) experiment~\cite{Erickcek2007}, and a balloon-based experiment (RRS)~\cite{Rich1987}. The solid grey line denotes the critical reference cross-section ($\bar{\sigma}_{e\mathrm{,ref,crit}}$).}
\end{figure}

From Eq.~\ref{Eqn:RefXS}, MAPP-1's exclusion limits for mCPs can be projected to the mC-SIDM scenario. Our results are presented in Fig.~\ref{fig:fig7}. Under the assumption of a $0.4\%$ minicharged subcomponent of dark matter, several additional exclusion regions are featured, including limits set by terrestrial direct-detection experiments~\cite{Mahdawi2018,Emken2019}, the X-ray quantum calorimetry (XQC) rocket experiment~\cite{Erickcek2007}, and the balloon-based experiment conducted by Rich et al.~\cite{Rich1987} (denoted as RRS). Also shown is the critical reference cross-section ($\bar{\sigma}_{e\mathrm{,ref,crit}}$)~\cite{Mahdawi2018,Emken2018,Emken2019} denoting the upper limit in the sensitivity of direct-detection experiments.

Notably, for both Run~3 and the HL-LHC, MAPP-1 can probe significant unconstrained regions of the mC-SIDM parameter space that are out of reach to direct-detection experiments. Moreover, as the authors of Ref.~\cite{Foroughi2021} pointed out, the exclusion limits imposed by accelerator-based searches do not have a dependence on the total minicharged dark matter subcomponent, $f_{\chi}$.

% ***** Conclusions ******
\section{\label{sec:Conc}Conclusions}
This study reported the latest results of our detailed investigation of the MoEDAL's Apparatus for Penetrating Particles (MAPP) Experiment's flagship benchmark scenario involving minicharged particles (mCPs). We considered a host of production mechanisms of mCPs in $pp$ collisions, including the Drell--Yan process and various meson decays. Our findings revealed highly competitive exclusion limits for the Phase-1 MAPP detector (MAPP-1) for mCPs produced at the LHC's Run~3 and the HL-LHC, which surpass the most stringent current bounds by over an order of magnitude across a sizable range of mCP masses. These results indicate that MAPP-1 demonstrates sensitivity to a wide range of unexplored parameter space and can detect effective charges as low as $8 \times 10^{-4}\:e$ and $6 \times 10^{-4}\:e$ for Run~3 and the HL-LHC, respectively. Additionally, in the minicharged strongly interacting dark matter scenario, MAPP-1 exhibits significant sensitivity to an unexplored and unique window of parameter space that is inaccessible to direct-detection experiments. These results and their associated analyses serve as the foundation for the future of mCP searches at the MAPP Experiment.

Several upgrades to the MAPP-1 detector are planned. First, an auxiliary detector for the MAPP-1 bar detector, known as the outrigger detector, aimed at improving the sensitivity of the experiment to high-mass, intermediate-charge mCPs, is currently under review by the LHC Experiments Committee. Second, multiple possible upgrades to the MAPP-1 detector to improve the overall sensitivity of the experiment for the HL-LHC are in discussion. 

We are excited to share that MAPP-1 is on track to begin data-taking at the LHC's Run~3 early into 2024. Furthermore, a complete study involving a comprehensive analysis of background sources and detailed mCP energy loss simulations is currently underway; additional production mechanisms, such as proton bremsstrahlung, are also actively under investigation. We anticipate the completion of this work during the Run-3 data-taking phase.

% ***** Acknowledgments ******    
\begin{acknowledgments}
We are grateful to the Natural Sciences and Engineering Research Council of Canada (NSERC) for partial financial support: Discovery Grant, SAPPJ-2019-00040; Research Tools and Instruments Grants, SAPEQ-2020-00001 and SAPEQ-2022-00005. M.d.M. thanks NSERC for partial financial support under Discovery Grant No. RGPIN-2016-04309. M.S. acknowledges support by the Generalitat Valenciana (GV) via the APOSTD Grant No. CIAPOS/2021/88. V.A.M. and M.S. acknowledge support by the GV via Excellence Grant No. CIPROM/2021/073, as well as by the Spanish MICIU/AEI and the European Union/FEDER via the Grant PID2021-122134NB-C21. V.A.M. was supported in part by the Ministry of Universities (Spain) via the Mobility Grant PRX22/00633. We also acknowledge with thanks the authors and maintainers of the \textsc{FeynRules}, \textsc{MadGraph}~5, \textsc{Pythia}~8, and \textsc{Geant}4 software packages. \\
\end{acknowledgments}

% ***** Other Back Matter ******
\paragraph*{Author Contributions:}

Conceptualization, J.P. and M.S.; methodology, M.d.M., A.M., P.-P.A.O., and M.S.; validation, A.M., A.S., and M.S.; formal analysis, A.M. (\textsc{Geant}4) and M.S. (\textsc{MadGraph}~5, \textsc{Pythia}~8); investigation, M.S.; resources, V.A.M. and J.P.; data curation, M.S.; writing---original draft preparation, P.-P.A.O. and M.S.; writing---review and editing, V.A.M., M.d.M., J.P., and M.S.; visualization, J.P. and M.S.; supervision, V.A.M. and J.P.; project administration, J.P.; funding acquisition, V.A.M., M.d.M., J.P., and M.S. All authors have read and agreed to the published version of the manuscript.

\bibliography{mCPs_at_MAPP}% Produces the bibliography via BibTeX.

\end{document}